\documentclass[graybox]{svmult}

\usepackage{mathptmx}       
\usepackage{helvet}         
\usepackage{courier}        
\usepackage{type1cm}        
\usepackage{makeidx}         
\usepackage{graphicx}        
\usepackage{multicol}        
\usepackage[bottom]{footmisc}
\usepackage{multirow}
\usepackage{amssymb}

\makeindex             

\begin{document}

\title*{Complex asteroseismology of the Slowly Pulsating B-type star HD74560}
\author{Walczak, P., Szewczuk, W. \& Daszy\'nska-Daszkiewicz, J.}
\institute{Szewczuk W., Walczak P., Daszy\'nska-Daszkiewicz J. \at Instytut Astronomiczny, Uniwersytet Wroc{\l}awski,
Kopernika 11, 51-622 Wroc{\l}aw, Poland, \email{szewczuk@astro.uni.wroc.pl}, \email{walczak@astro.uni.wroc.pl},\email{daszynska@astro.uni.wroc.pl}}
%
%
\maketitle
\abstract*{We present the results of complex seismic modelling of the Slowly Pulsating B-type star HD74560. The star pulsates in five frequencies detected in photometric observations. For all these frequencies, we identify the mode degree, $\ell$. For two of them, found also in spectroscopic data, we are able to derived the empirical values of the complex nonadiabatic parameter $f$. We test effects of chemical composition and opacity data. Our results show that the properties of seismic models of SPB stars differ significantly from those of the more massive $\beta$ Cephei stars.}

\abstract{We present the results of complex seismic modelling of the Slowly Pulsating B-type star HD74560. The star pulsates in five frequencies detected in photometric observations. For all these frequencies, we identify the mode degree, $\ell$. For two of them, found also in spectroscopic data, we are able to derived the empirical values of the complex nonadiabatic parameter $f$. We test effects of chemical composition and opacity data. Our results show that the properties of seismic models of SPB stars differ significantly from those of the more massive $\beta$ Cephei stars.}

\section{Introduction}
\label{sec:1}
The Slowly Pulsating B-type stars (SPB) are very interesting targets for asteroseismic analysis. They pulsate in high-order g-modes, which penetrate the deep interior of the star. Because of it, these types of modes are extremely useful for studying the overshooting efficiency. This effect is poorly understood and analysing pulsations of the SPB stars is a unique opportunity to get constraints on its range.

HD74560 is a bright object with the parallax of $\pi=6.73(17)$ mas. \cite{W1991} discovered line profile variations with two frequencies
and classified the star as a SPB variable. He also found two candidates for a third frequency but it was impossible to determine
the physical one owing to the aliasing. \cite{CA2002} found at least five frequencies in the Geneva photometry.
Only the dominant frequency, $\nu_1$, agrees with the results of \cite{W1991}.
\cite{CA2002} found also two of their five frequencies in the radial velocity variations.

In addition to the fitting frequencies our seismic analysis makes use of the complex nonadiabatic $f$-parameter,
defined as the ratio of the bolometric flux changes to the radial displacement (see i.e. \cite{DDDP2003, DDW2009}).
Fitting frequencies and the corresponding values of $f$ is termed {\it complex asteroseismology}
and it can provide stronger constraints on the model parameters and theory.

The paper is organized as follows. Results of mode identification are shown in Section\,\ref{sec:2}.
In Section\,\ref{sec:3} we present our attempt towards complex seismic modeling of the SPB star HD74560. Conclusions end the paper.
\section{Mode identification}
\label{sec:2}

The five frequencies of HD74560 are shown in the first column of Table\,\ref{tab:1}.
These values were taken from \cite{CA2002}. The frequencies range from about 0.4 up to 0.6 c/d which is typical for the SPB stars.

In order to identify the degree, $\ell$, of pulsational modes we incorporated the theoretical and empirical values of the $f$-parameter. In the first case we compared the theoretical amplitude ratios and phase differences in different passbands with observational counterparts. The appropriate
theoretical values of $f$ result from linear nonadiabatic calculations of stellar pulsations. We used the seven Geneva passbands and the non-LTE models of stellar atmospheres by \cite{LH2007}. The identified mode degrees are presented in the third column of Table\,\ref{tab:1}. As we can see, the most probable identification for all frequencies are the dipole modes, $\ell=1$. The $\nu_2$ and $\nu_3$ frequencies can be also the quadruple modes, $\ell=2$.

In the second method, we derived the mode degrees together with the empirical values of the $f$-parameter.
We have to stress that in the case of the B-type pulsators, we are able to extract the empirical $f$ only for modes which are seen both in photometry and spectroscopy (\cite{DDDP2005}). Therefore, by means of this method we identified the mode degrees only for two frequencies, $\nu_1$ and $\nu_2$. The results, given in the last column of Table\,\ref{tab:1}, are in agreement with mode degrees derived with the previous method.
The only difference is for $\nu_1$; it can be the quadruple mode as well.

Summarizing, all pulsational frequencies of HD74560 can be the dipole modes, while $\nu_2$ and $\nu_3$ can also be the quadruple modes.

\begin{table}
\begin{center}
 \caption{Pulsational frequencies of HD74560 with corresponding values of the Geneva $V$ amplitudes
  and the most probable mode identification derived with the theoretical and empirical values of the $f$-parameter.}
 \label{tab:1}
 \begin{tabular}{lccc}
\hline\noalign{\smallskip}
 \multirow{2}{2.9cm}{Frequency [c/d]}& the Geneva $V$ amplitude& Mode degree, $\ell$,& Mode degree, $\ell$,\\
                        &  [mmag]      & theoretical $f$-parameter &empirical $f$-parameter\\
\noalign{\smallskip}\svhline\noalign{\smallskip}
    $\nu_1$=0.64472 & 14.4(4)&1   &1,2\\
    $\nu_2$=0.39578 & 4.3(4) &1,2 &1,2\\
    $\nu_3$=0.44763 & 3.2(4) &1,2 &-  \\
    $\nu_4$=0.82281 & 3.7(4) &1   &-  \\
    $\nu_5$=0.63567 & 3.0(4) &1   &-  \\
\noalign{\smallskip}\hline\noalign{\smallskip}
 \end{tabular}
 \end{center}
\end{table}

\section{Complex asteroseismology}
\label{sec:3}
In the previous section we determined the mode degrees, $\ell$, of the observed frequencies of HD74560. In the next step, we constructed models reproducing the frequencies. Firstly, we chose two modes, $\nu_1$ and $\nu_4$. The $\nu_1$ frequency is the strongest, well identified mode, and the $\nu_4$ frequency was chosen because it has the highest value. Relatively high values of frequencies facilitate modelling. Apart from that, $\nu_2$, $\nu_3$, $\nu_4$ and $\nu_5$ have rather similar amplitudes (see the second column of Table\,\ref{tab:1}).

In order to construct seismic models we have to know the azimuthal number, $m$, and radial order, $n$, of our modes. We assumed that pulsational modes of HD74560 are axisymmetric. The star is rather a slow rotator, ($V\sin{i}=35$ km/s according to \cite{GS2000}), and the results should not differ significantly from those presented here.

The frequency spectrum of high-order g-modes is very dense and identification of the radial order is difficult, but it turned out that fitting two frequencies constrains $n$ quite strongly. We were able to find seismic models only for two combinations of the radial orders. The first possibility is: $\nu_1$ is the $\ell=1$, g$_{22}$ mode and $\nu_4$ is the $\ell=1$ g$_{17}$ mode. The second one is:  $\nu_1$ is the $\ell=1$ g$_{17}$ mode and $\nu_4$ is the $\ell=1$ g$_{13}$ mode.

\begin{figure}[h]
\includegraphics[clip,width=5.8cm,height=5.3cm]{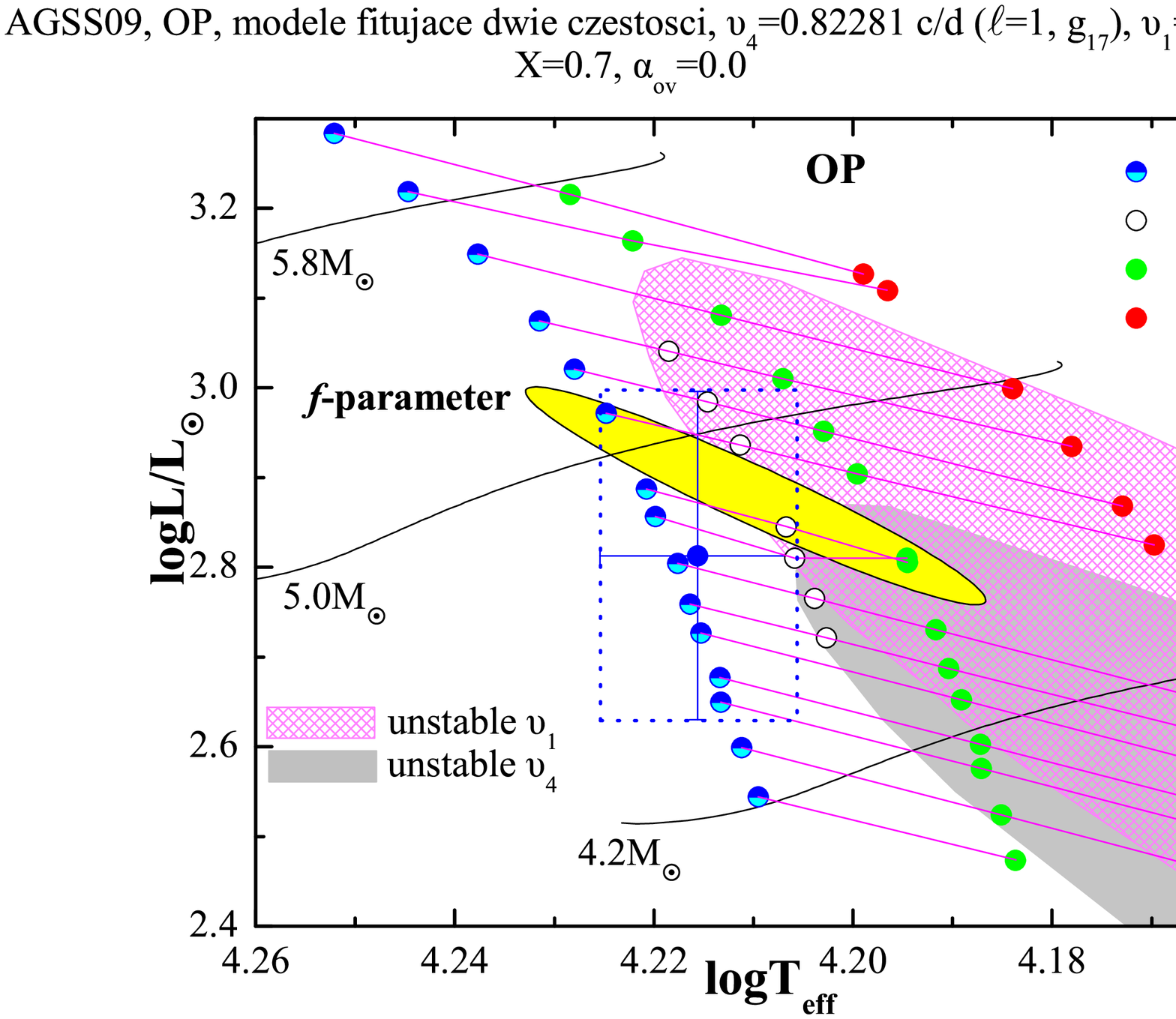}
\includegraphics[clip,width=5.8cm,height=5.3cm]{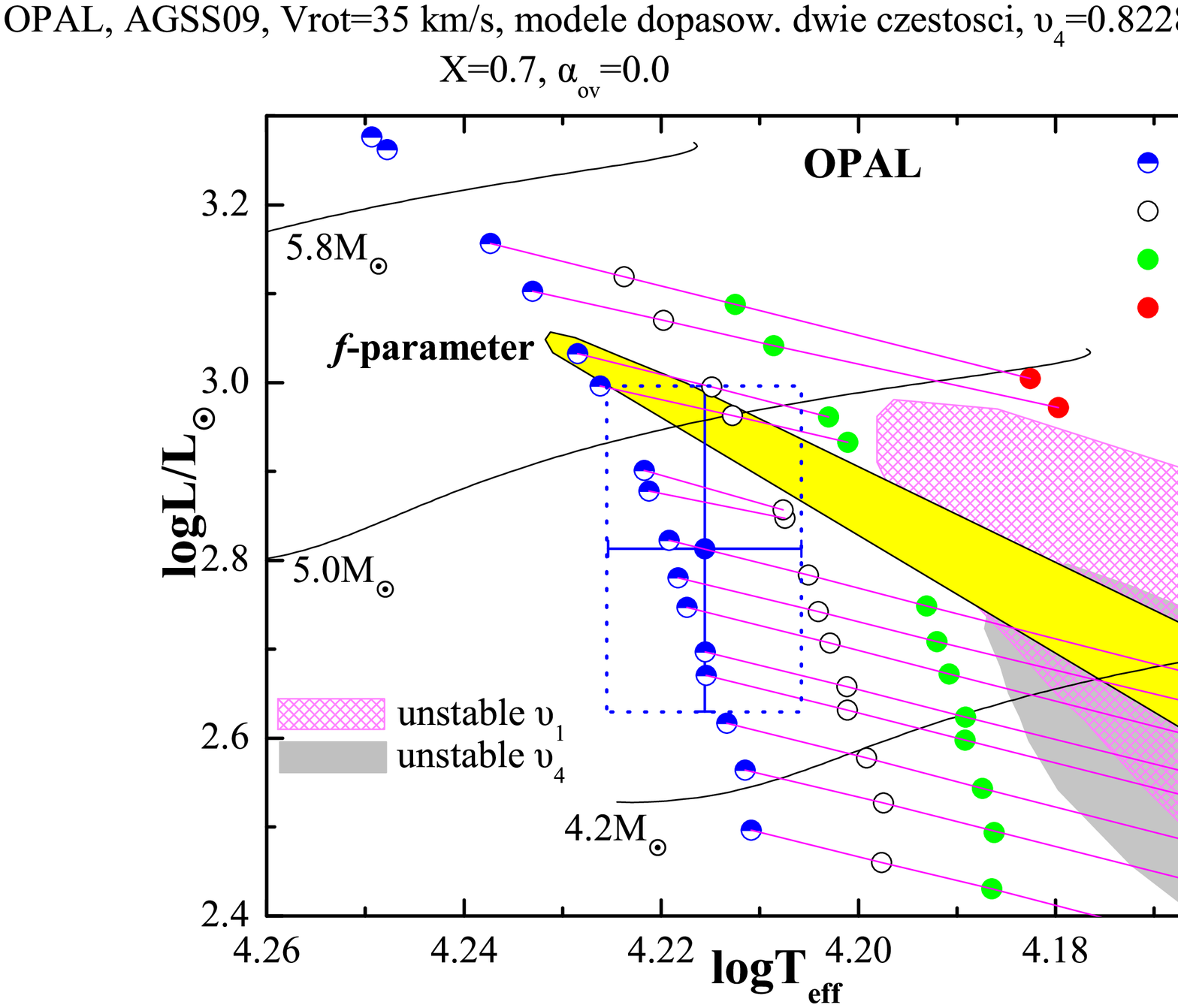}
\caption{Seismic models of the SPB star HD74560 fitting two frequencies, $\nu_1$ and $\nu_4$, on the HR diagram. Results for four values of metallicity, $Z$, are shown. The left and right panels show models obtained with the OP and OPAL data, respectively.
We marked also regions where the $\nu_1$ and $\nu_4$ modes are pulsational unstable.}
\label{fig:1}       
\end{figure}

In Fig.\,\ref{fig:1}, we show the position of seismic models of HD74560 on the HR diagram for several values of the metallicity parameter, $Z$. These models fit the $\nu_1$ and $\nu_4$ frequencies assuming the first combination of the radial orders. We stress, that for a given $Z$ there are no seismic models between points shown in Fig\,\ref{fig:1}. The error box was taken from \cite{CA2002} and Pamyatnykh (private communication). The left and right panels show the results for the OP and OPAL opacity tables, respectively. All models are computed without any overshooting of the convective core. The hatched and grey areas indicate the instability regions of the $\nu_1$ and $\nu_4$ modes, respectively. The areas labeled as "$f$-parameter" indicate models fitting the empirical value of $f$ for the $\nu_1$ mode.

As we can see, the effective temperature of seismic models strongly depends on metallicity. Both with the OP and OPAL opacities, we need quite low value of $Z$ to catch models inside the observational error box ($Z<0.007$). The instability regions computed with different opacities differ significantly. In the case of the OPAL tables we need much higher value of $Z$ to excite the $\nu_1$ and $\nu_4$ modes ($Z>0.007$). With the OP data, models with metallicity of $Z=0.006$ are already unstable.


Relying on the $f$ parameter, it is difficult to judge which opacity data are more suitable. Although with the OPAL tables, the $f$-parameter area is larger, most of the models fitting $f$ are located outside of the error box.

We also performed seismic analysis assuming $\alpha_{\rm{ov}}=0.1$, 0.2 and 0.4. In these cases, we did not find any model fitting the empirical $f$-parameter within
$1\sigma$ errors. Thus, the $f$-parameter can be used to constrain the overshooting range.

Interesting results were obtained with the second combination of radial orders: $\nu_1$ is the $\ell=1$ g$_{17}$ mode and $\nu_4$ is the $\ell=1$, g$_{13}$ mode. In this case, seismic models located inside the observational error box have much higher metallicity, $Z\approx0.014$. As a consequence, all models fitting the observational effective temperature and luminosity have unstable modes $\nu_1$ and $\nu_4$ . Moreover, inside of the error box there is at least one model fitting the empirical value of the $f$-parameter for the $\nu_1$ mode.

\section{Conclusions}

Seismic modelling of the SPB stars is quite challenging. There are seismic models inside the error box computed with the OP and OPAL opacities and with different combinations of radial orders. Some of these models reproduce also the empirical value of the $f$-parameter, but to get stronger constraints on stellar parameters, we need to fit more than two frequencies.

For the time being, we can conclude that there are some preferences towards the OP data owing to the larger instability areas for $\nu_1$ and $\nu_4$. All seismic models preferred small overshooting parameter, $\alpha_{\rm{ov}}\lesssim0.1$. Interesting constraints were derived for metallicity.  The values of $Z$ can be either very small, $\sim0.005-0.007$ (assuming the first combination of the radial orders), or quite high $\sim0.013-0.015$ (for the second combination of $n$).
The other values of $Z$ are excluded.
\\
\\
\textbf{Acknowledgments} The authors acknowledge partial financial support from the Polish
MNiSW grant No. N N203 379 636.


\begin{thebibliography}{99.}%

\bibitem{DDDP2003} Daszy\'nska-Daszkiewicz J., Dziembowski W.A., Pamyatnykh A.A.: A\&A, \textbf{407}, 999 (2003)
\bibitem{DDDP2005} Daszy\'nska-Daszkiewicz J., Dziembowski W.A., Pamyatnykh A.A.: A\&A, \textbf{441}, 641 (2005)
\bibitem{DDW2009} Daszy\'nska-Daszkiewicz J., Walczak, P.: MNRAS, \textbf{398}, 1961, (2009)
\bibitem{CA2002} De Cat,P., Aerts, C.: A\&A \textbf{393}, 965 (2002)
\bibitem{GS2000} G{\l}ebocki, R., Stawikowski, A.: AcA, \textbf{50}, 509 (2000)
\bibitem{LH2007} Lanz T., Hubeny I.: ApJS, \textbf{169}, 83 (2007)
\bibitem{W1991}Waelkens, C.:, A\&A, \textbf{246}, 453 (1991)

\bigskip

\end{thebibliography}
\end{document}